\documentclass[12pt]{article}
\usepackage{graphicx}
\usepackage{lscape}
\usepackage{citesort}

\begin{document}

\def\ds{\displaystyle}
\def\beq{\begin{equation}}
\def\eeq{\end{equation}}
\def\bea{\begin{eqnarray}}
\def\eea{\end{eqnarray}}
\def\beeq{\begin{eqnarray}}
\def\eeeq{\end{eqnarray}}
\def\ve{\vert}
\def\vel{\left|}
\def\ver{\right|}
\def\nnb{\nonumber}
\def\ga{\left(}
\def\dr{\right)}
\def\aga{\left\{}
\def\adr{\right\}}
\def\lla{\left<}
\def\rra{\right>}
\def\rar{\rightarrow}
\def\nnb{\nonumber}
\def\la{\langle}
\def\ra{\rangle}
\def\ba{\begin{array}}
\def\ea{\end{array}}
\def\tr{\mbox{Tr}}
\def\ssp{{\Sigma^{*+}}}
\def\sso{{\Sigma^{*0}}}
\def\ssm{{\Sigma^{*-}}}
\def\xis0{{\Xi^{*0}}}
\def\xism{{\Xi^{*-}}}
\def\qs{\la \bar s s \ra}
\def\qu{\la \bar u u \ra}
\def\qd{\la \bar d d \ra}
\def\qq{\la \bar q q \ra}
\def\gGgG{\la g^2 G^2 \ra}
\def\q{\gamma_5 \not\!q}
\def\x{\gamma_5 \not\!x}
\def\g5{\gamma_5}
\def\sb{S_Q^{cf}}
\def\sd{S_d^{be}}
\def\su{S_u^{ad}}
\def\ss{S_s^{??}}
\def\sbp{{S}_Q^{'cf}}
\def\sdp{{S}_d^{'be}}
\def\sup{{S}_u^{'ad}}
\def\ssp{{S}_s^{'??}}
\def\sig{\sigma_{\mu \nu} \gamma_5 p^\mu q^\nu}
\def\fo{f_0(\frac{s_0}{M^2})}
\def\ffi{f_1(\frac{s_0}{M^2})}
\def\fii{f_2(\frac{s_0}{M^2})}
\def\O{{\cal O}}
\def\sl{{\Sigma^0 \Lambda}}
\def\es{\!\!\! &=& \!\!\!}
\def\ap{\!\!\! &\approx& \!\!\!}
\def\ar{&+& \!\!\!}
\def\ek{&-& \!\!\!}
\def\kek{\!\!\!&-& \!\!\!}
\def\cp{&\times& \!\!\!}
\def\se{\!\!\! &\simeq& \!\!\!}
\def\eqv{&\equiv& \!\!\!}
\def\kpm{&\pm& \!\!\!}
\def\kmp{&\mp& \!\!\!}


\def\simlt{\stackrel{<}{{}_\sim}}
\def\simgt{\stackrel{>}{{}_\sim}}


\title{
         {\Large
                 {\bf
Analysis of the semileptonic $B_c \rar B_u^\ast \ell^+ \ell^-$ decay from
QCD sum rules
                 }
         }
      }

\author{\vspace{1cm}\\
{\small T. M. Aliev \thanks
{e-mail: taliev@metu.edu.tr}~\footnote{permanent address:Institute
of Physics,Baku,Azerbaijan}\,\,,
M. Savc{\i} \thanks
{e-mail: savci@metu.edu.tr}} \\
{\small Physics Department, Middle East Technical University,
06531 Ankara, Turkey} }

\date{}

\begin{titlepage}
\maketitle
\thispagestyle{empty}

\begin{abstract}
We analyze the semileptonic $B_c \rar B_u^\ast \ell^+ \ell^-$ decay in the
frame work of the Standard Model. We calculate the $B_c$ to $B_u^\ast$
transition form factors in QCD sum rules. Analytical
expressions for the spectral densities and gluon condensates are presented.
The branching ratio of the $B_c \rar B_u^\ast \ell^+ \ell^-$ decay is
calculated, and it is obtained that this decay can be detectable at
forthcoming LHC machines.
\end{abstract}

~~~PACS numbers: 12.60.--i, 13.30.--a. 13.88.+e
\end{titlepage}

\section{Introduction}

The double heavy $B_c$ mesons have been firstly discovered by the CDF
Collaboration \cite{R7301}, with mass $m_{B_c}^{exp} = (6.4 \pm 0.39 \pm
0.13)~GeV$. Recently CDF Collaboration \cite{R7302} announced an accurate
determination of the $B_c$ meson mass, $m_{B_c} = (6.2857 \pm 0.0053 \pm
0.0012)~GeV$. The study of the $B_c$ meson has received a great interest,
due to its special properties: firstly, its lowest bound state is composed of
two heavy (charm and beauty) quarks with open flavor; secondly, this
meson attracts the interest of physicists for checking predictions of the
pertubative QCD in the laboratory; and lastly, the weak decay channels of $B_c$ 
meson are reacher compared to the corresponding decay channels of $B_q(q=u,d,s)$ 
and can be divided into three classes:
\begin{itemize}
\item $b$ quark decaying weakly, with the $c$ quark as spectator, e.g., 
$B_c \rar J/\psi \ell \bar{\nu}_\ell$;

\item  $c$ quark decaying weakly, with the $b$ quark as spectator, e.g., 
$B_c \rar B_s \ell \bar{\nu}_\ell$;
   
\item The annihilation channels like $B_c \rar \ell \bar{\nu}_\ell$,
$B_c \rar D_s^{\ast-} \bar{K}^{0\ast}$.
\end{itemize}

There are quite a few number of theoretical works studying various leptonic,
semileptonic and hadronic exclusive decay channels of $B_c$ meson (for a
review, see \cite{R7303}).

From experimental point of view, the study of weak semileptonic decays of
$B_c$ meson is quite important for the determination of
Cabibbo--Kobayashi--Maskawa (CKM) matrix elements, leptonic decay constant
$f_{B_c}$, etc.

Much more $B_c$ mesons and more detailed information about their decay
properties are expected at the forthcoming LHC accelerator. In particular,
this holds true for the dedicated detectors BTeV and LHCB which are
specially designed for the analysis of $B$ physics where one expects to see
up to $10^{10}~B_c$ mesons per year with a luminosity
$10^{34}~cm^{-2}s^{-1}$ \cite{R7304}
 
The rare decays constitute on of the most important class of decays for a
more precise determination of the parameters of the SM, as well as looking
new physics beyond the SM \cite{R7305,R7306,R7307}, since the flavor
changing neutral currents (FCNC) are absent in the SM at tree level and
appear only at loop level, due to the running of virtual particle in the
loop.

Of the FCNC processes involving $K$ and $B$ mesons, main attention has been 
focused on $B^0-\bar{B}^0$ mixing, $b \rar s \ell^+ \ell^-$, $b \rar s \gamma$, 
$s \rar d \ell^+ \ell^-$, $s \rar d \nu \bar{\nu}$, etc, since top quark
runs in the loop.

Study of the FCNC decays in charm sector has not received enough attention.
This can be explained by the fact that in the SM, $D^0-\bar{D}^0$ mixing
\cite{R7308,R7309}, as well as FCNC decays \cite{R7310,R7311,R7312}, are
expected to be very small. Moreover, long distance effects are quite huge,
since the loop in charm decay involves light down quarks.

In the present work we present a detailed analysis of the semileptonic $B_c
\rar B_u^\ast \ell^+ \ell^-$ decay in the framework of sum rules method.
This mode is likely to be observed in forthcoming accelerator experiments.
Note that the radiative decay $B_c \rar B_u^\ast \gamma$, which at quark level 
described by $c \rar u + \gamma$ transition is investigated in
\cite{R7313,R7314}.

The plan of this work is as follows: In the following section QCD sum rules
of the three--point correlators are considered and sum rules for the form
factors that are responsible for the $B_c \rar B_u^\ast \ell^+ \ell^-$ decay
are constructed. In section 3 we present our numerical results and
conclusions.

\section{Sum rules for transition form factors}

In the present section we derive sum rules for the form factors that
controls the $B_c \rar B_u^\ast \ell^+ \ell^-$ decay This decay is described
by the $c \rar u \ell^+ \ell^-$ transition at quark level. The matrix
element for the $c \rar u \ell^+ \ell^-$ decay can be written in the
following form:
\bea
\label{e7301}   
{\cal M} \es \frac{G \alpha}{4 \sqrt{2} \pi} \Bigg[ C_9^{eff}(m_c) \bar{u}
\gamma_\mu (1-\gamma_5) c \bar{\ell} \gamma_\mu \ell + 
C_{10}(m_c) \bar{u} \gamma_\mu (1-\gamma_5) c \bar{\ell} \gamma_\mu
\gamma_5 \ell \nnb \\
\ek 2 m_c C_7(m_c) \bar{u}i \sigma_{\mu\nu} \frac{q^\nu}{q^2}
(1+\gamma_5) c \bar{\ell} \gamma_\mu \ell \Bigg]~.
\eea
At $\mu=m_c$, $C_9^{eff}$ is given by \cite{R7315}
\bea
\label{e7302}
C_9^{eff}(m_c) \es C_9(m_W) + \sum_{i=d,s} \lambda_i \left[
-\frac{2}{9} \ln \frac{m_i^2}{m_W^2} + \frac{8}{9} \frac{z_i^2}{\bar{s}} 
-\frac{1}{9} \ga 2 + \frac{4 z_i^2}{\bar{s}} \dr \sqrt{ \vel 1 -
\frac{4 z_i^2}{\bar{s}} \ver } T(z_i) \right]~,
\eea
where 
\bea
T(z_i) = \left\{ \begin{array}{lc}
2 \arctan \left[\frac{1}{\sqrt{4 z_i^2/\bar{s} -1}} \right]& \ga \mbox{\rm for}
~\bar{s} < 4 z_i^2 \dr ~,\\ \\
\ln \vel \frac{1+\sqrt{1-4 z_i^2/\bar{s}}}{1-\sqrt{1-4 z_i^2/\bar{s}}} \ver - i \pi & 
\ga \mbox{\rm for}~ \bar{s} > 4 z_i^2 \dr
~.\end{array} \right. \nnb~,
\eea
where $\bar{s} = s/m_c^2$, $z_i = m_i/m_c$. The logarithmic dependence on
the internal quark mass $m_i$ in $T(z_i)$ in Eq. (\ref{e7302})
cancels a similar term in the function $F_1(x_i)$ entering in $C_9(m_W)$,
leaving no spurious divergences in the limit $m_i \rar 0$. 
It should be noted here that QCD corrections do not effect $C_{10}$, i.e., 
$C_{10}(m_c) = C_{10}(m_W)$.

The QCD corrections are particularly important for the Wilson coefficient $C_7$
and in the numerical analysis we use the two--loop QCD--corrected value of
$C_7^{eff}$ which is calculated in \cite{R7311}.

The values of the Wilson coefficients at $\mu = m_W$ are given by the
following expressions (see for example \cite{R7310,R7313})
\bea
\label{e7303}
C_7(m_W) \es - \sum_{i=d,s,b} \lambda_i F_{2}(x_i)~, \nnb \\
C_9(m_W) \es \frac{1}{s_W^2} \sum_{i=d,s,b} \lambda_i \left[ 
\ga C^{box}(x_i) + C^{Z}(x_i) \dr -
2 s_W^2 \ga F_{1}(x_i) + C^Z(x_i) \dr \right]~,\nnb \\
C_{10}(m_W) \es -\frac{1}{s_W^2} \sum_{i=d,s,b} \lambda_i \ga
C^{box}(x_i) + C^Z(x_i) \dr~,
\eea
where $\lambda_i = V_{ci} V_{ui}^\ast$, $x_i = m_i^2/m_W^2$. The functions
$F_1(x)$, $F_2(x)$, $C^{box}(x_i)$ and $C^Z(x_i)$ are those
derived in \cite{R7316} and are all given in Appendix--A. 

Similar to the $b \rar s \ell^+ \ell^-$ transition, the Wilson coefficient in
the $c \rar u \ell^+ \ell^-$ transition receives long distance contributions
which have their origin in the real $\bar{q}q$ intermediate state, i.e.,
$\rho$ and $\omega$ mesons. These contributions can be written via
the following replacement in $C_9^{eff}(m_c)$ \cite{R7317}
\bea
C_9^{eff} \rar C_9^{eff}  + \frac{3 \pi}{\alpha^2} 
\sum_i \kappa_{i} \ds{\frac{ m_{V_i} \Gamma_{V_i \rar \ell^+ \ell^-}}
{m_{V_i}^2 - \hat{s} - i m_{V_i} \Gamma_{V_i}} }~, \nnb
\eea
where $m_{V_i}$ and $\Gamma_{V_i}$ are the resonance mass and width. The
Fudge factor $\kappa_{i}$ is determined in \cite{R7310} to have the values
$\kappa_{\rho}=0.7$ and $\kappa_{\omega}=3.1$. 

Having the matrix element for $c \rar u \ell^+ \ell^-$ transition at hand, 
our next problem is the calculation of the matrix element for the 
$B_c \rar B_u^\ast \ell^+ \ell^-$ decay. It follows from Eq. (\ref{e7301})
that, in order to calculate the amplitude of the $B_c \rar B_u^\ast \ell^+
\ell^-$ decay, the following matrix elements are needed
\bea
&&\lla B_u^\ast \vel \bar{u} \gamma_\mu (1-\gamma_5) c \ver B_c \rra~, \nnb \\ 
&&\lla B_u^\ast \vel \bar{u} i \sigma_{\mu\nu} q^\nu (1+\gamma_5) b \ver B_c
\nnb 
\rra~.
\eea
These hadronic matrix elements of the $B_c \rar B_u^\ast \ell^+ \ell^-$
decay can be parametrized in terms of the form factors in the following way:
\bea
\label{e7304}
\lla B_u^\ast(p^\prime,\varepsilon) \vel \bar{u} \gamma_\mu (1-\gamma_5) c
\ver B_c (p)\rra \es \epsilon_{\mu\nu\alpha\beta}
\varepsilon^{\ast\nu} p^\alpha p^{\prime\beta} \frac{2 V(q^2)}{m_{B_c} +
m_{B_u^\ast}} \nnb \\
\ek i \Bigg[ \varepsilon_\mu^\ast (m_{B_c} + m_{B_u^\ast}) A_1(q^2) 
-(\varepsilon^\ast q) {\cal P}_\mu \frac{A_2(q^2)}{m_{B_c} +
m_{B_u^\ast}}\nnb \\ 
\ek (\varepsilon^\ast q) \frac{2 m_{B_u^\ast}}{q^2} [A_3(q^2)
-A_0(q^2)] q_\mu \Bigg]~, \\ \nnb \\
\label{e7305}
\lla B_u^\ast(p^\prime,\varepsilon) \vel \bar{u} \sigma_{\mu\nu} q^\nu (1+\gamma_5) c  
\ver B_c (p)\rra \es 2 i \epsilon_{\mu\rho\alpha\beta}
\varepsilon^{\ast\rho} p^\alpha p^{\prime\beta} T_1(q^2) +
[\varepsilon_\mu^\ast (m_{B_c}^2 - m_{B_u^\ast}^2) 
- (\varepsilon^\ast q) {\cal P}_\mu ] T_2(q^2) \nnb \\
\ar (\varepsilon^\ast q) \Bigg[ q_\mu - \frac{q^2}{m_{B_c}^2 - m_{B_u^\ast}^2}
{\cal P}_\mu \Bigg] T_3(q^2)~,
\eea
where ${\cal P}_\mu = (p+p^\prime)_\mu$, $q_\mu=(p-p^\prime)_\mu$ and
$\varepsilon^\ast$ is the polarization 4--vector of the vector $B_u^\ast$
meson. Note that the form factor $A_3(q^2)$ can be written as a linear
combination of $A_1$ and $A_2$ as:
\bea
\label{e7306}
A_3(q^2) = \frac{m_{B_c} + m_{B_u^\ast}}{2 m_{B_u^\ast}} A_1(q^2) - \frac{m_{B_c} -
m_{B_u^\ast}}{2 m_{B_u^\ast}} A_2(q^2)~,
\eea
and in order to guarantee the finiteness of the results at $q^2=0$,
$A_3(0)=A_0(0)$ should be satisfied.
 
The identity
\bea
\sigma_{\mu\nu}\gamma_5 = - \frac{i}{2} \epsilon_{\mu\nu\alpha\beta}
\sigma^{\alpha\beta}~,\nnb
\eea
leads to the following relation between the form factors $T_1$ and $T_2$,
\bea
T_1(0) = T_2(0)~.\nnb
\eea
Moreover, it is shown in \cite{R7318} that the form factors $T_1(q^2)$ and
$A_0(q^2)$
are also related, i.e.,
\bea
A_0(q^2) = -T_1(q^2)~,\nnb
\eea  
and so are $T_3(q^2)$, $A_1(q^2)$ and $A_2(q^2)$ through the relation
\bea
\label{e7307}
T_3(q^2) = - \Bigg[\frac{m_{B_c} + m_{B_u^\ast}}{2 m_{B_u^\ast}} A_1(q^2) -
\frac{m_{B_c}^2 + 3 m_{B_u^\ast}^2 - q^2}{2 m_{B_u^\ast}(m_{B_c} +
m_{B_u^\ast})} A_2(q^2)\Bigg]~.
\eea
It follows from these relations that in describing the $B_c \rar B_u^\ast
\ell^+ \ell^-$ decay, we need to know only $V$, $A_1$ and  $A_2$. In
further analysis the values of these form factors at $q^2=0$ are needed
(see section--3). Note that, at this point the form factor $T_1(0)$ is 
calculated in \cite{R7314}, and hence we do not present its expression in
this work. 
   
As has already been noted, in order to calculate the these form factors
appearing in the $B_c \rar B_u^\ast \ell^+ \ell^-$ decay, we explore the
three--point QCD sum rules \cite{R7319}.

For the evolution of the above--mentioned form factors in the frame work of
the QCD sum rules, we start with the following three-point correlation
functions: 
\bea
\label{e7308}
\Pi_{\nu\mu} = - \int d^4x d^4y e^{-i (p_1 x + p_2 y)} 
\la 0 \ve {\cal T} \left\{ J_{B_u^\ast \nu}(y) J_\mu(0) J_{B_c}(x) 
\right\} \ve 0 \ra~, 
\eea
where $J_{B_u^\ast}^\nu=\bar{b} (y) \gamma^\nu u(y)$ is the interpolating 
current of the $B_u^\ast$ meson, $J_\mu=\bar{u}(0) \gamma_\mu (1+\gamma_5)
c(0)$, and $J_{B_c}=\bar{c}(x) i (1+\gamma_5) b(x)$ is the interpolating 
current of the $B_c$ meson. The contribution of the scalar part of this 
current to $B_c$ is zero.

The Lorentz structures of these correlators can be written in the following
forms:
\bea
\label{e7309}
\Pi_{\nu\mu}^{(V+A)} \es \epsilon_{\nu\mu\alpha\beta} \,p^\alpha
p^{\prime\beta} \Pi_V + \Pi_{A_1} g_{\mu\nu} + \Pi_2 {\cal P}_\mu {\cal
P}_\nu + \Pi_{A_2} {\cal P}_\nu q_\mu + \Pi_3 {\cal P}_\mu q_\nu + 
\Pi_4 q_\mu q_\nu~.
\eea 

The phenomenological part of these correlators can be calculated by
inserting a complete set of intermediate states with the same quantum
numbers the currents $J_{B_u^\ast}$ and $J_{B_c}$ posses in the correlation
functions (\ref{e7304}) and (\ref{e7305}), and expressing these functions as
the sum of the contributions of the lowest lying and excited states, we get
\bea
\label{e7310}
\Pi_{\nu\mu} (p_1^2,p_2^2,q^2) = - \frac{\la 0 \ve J_{B_u^\ast \nu} \ve
B_u^\ast \ra \la B_u^\ast \ve J_\mu \ve B_c \ra \la B_c \ve J_{B_c} \ve 0
\ra}
{(p_1^2 - m_{B_c}^2 ) ( p_2^2 - m_{B_u^\ast}^2)} 
+ \mbox{\rm contributions from higher states}~.
\eea
The matrix elements in (\ref{e7310}) are defined as follows:
\bea
\label{e7311}
\la 0 \ve J_{B_u^\ast}^\mu \ve B_u^\ast \ra \es f_{B_u^\ast} m_{B_u^\ast}
\varepsilon^{\ast\mu} \nnb \\
i \la B_c \ve \bar{c} (1+\gamma_5) b \ve 0 \ra \es \frac{f_{B_c}
m_{B_c}^2}{m_b+m_c}~.
\eea 

Performing summation over the polarization of $B_u^\ast$ meson 
on the matrix elements $\la B_u^\ast \ve \bar{u} \gamma_\mu (1+\gamma_5) c
\ve B_c \ra$ that are given in Eq. (\ref{e7304}), we get
for the physical part:
\bea
\label{e7312}
\Pi_{\nu\mu}^{(V+A)} \es - \frac{f_{B_c} f_{B_u^\ast} m_{B_c}^2}
{(m_b+m_c) (p_1^2 - m_{B_c}^2 ) ( p_2^2 - m_{B_u^\ast}^2)}
\Bigg\{ \epsilon_{\nu\mu\alpha\beta} p^\alpha p^{\prime\beta} \frac{2
V}{m_{B_c} + m_{B_u^\ast} } \nnb \\
\ar i (m_{B_c} + m_{B_u^\ast}) \Bigg( -g_{\mu\nu} + \frac{
({\cal P}-q)_\mu ({\cal P}-q)_\nu}{4 m_{B_u^\ast}^2} \Bigg) A_1 \nnb \\
\ek \frac{i}{m_{B_c} + m_{B_u^\ast}} {\cal P}_\mu \Bigg( - q_\nu + 
\frac{p^\prime q ({\cal P}-q)_\nu}{2 m_{B_u^\ast}^2} \Bigg) A_2 \nnb \\
\ek \frac{2 i m_{B_u^\ast}}{q^2} q_\mu \Bigg( - q_\nu +
\frac{p^\prime q ({\cal P}-q)_\nu}{2 m_{B_u^\ast}^2} \Bigg)  
(A_3 - A_0) \Bigg\}~.
\eea
The expressions of the form factors $V$, $A_1$ and $A_2$ can be determined
by choosing the coefficients of the Lorentz structures
$\epsilon_{\nu\mu\alpha\beta} p^\alpha p^{\prime\beta}$, $i g_{\mu\nu}$ and
${\cal P}_\mu q_\nu$ in the correlator function $\Pi_{\mu\nu}^{(V+A)}$,
respectively.

On the other side, the three--point correlator function can be calculated by
operator product expansion (OPE) in the deep Euclidean region $p_1^2 \ll
(m_b + m_c)^2$, $p_2^2 \ll m_b^2$.

The time ordered products of currents in the three--point correlator
function in Eq. (\ref{e7301}) can be expanded in terms of a series of local
operators with increasing dimension, as is shown in the following:
\bea
\label{e7313}
-\int d^4x d^4y e^{-i(p x - p^\prime y)} {\cal T} \Big\{J_{B_u^\ast\nu}
J_\mu J_{B_c} \Big\} \es
(C_0)_{\nu\mu} I + (C_3)_{\nu\mu} \bar{q} q + (C_4)_{\nu\mu} G_{\alpha\beta}
G^{\alpha\beta} \nnb \\
\ar (C_5)_{\nu\mu} \bar{q} \sigma_{\alpha\beta}
G^{\alpha\beta} q + (C_6)_{\nu\mu} \bar{q} \Gamma q \bar{q} \Gamma^\prime q~,
\eea  
where $(C_i)_{\nu\mu}$ are the Wilson coefficients, $I$ is the unit
operator, $G_{\alpha\beta}$ is the gluon field strength tensor, $\Gamma$ and
$\Gamma^\prime$ are the matrices appearing in the calculations. Considering
the vacuum expectation value of the OPE, the correlator function can be
written in terms of the local operators as:
\bea
\label{e7314}
\Pi_{\nu\mu} (p_1^2,p_2^2,q^2) \es (C_0)_{\nu\mu} + (C_3)_{\nu\mu} \la
\bar{q} q \ra + (C_4)_{\nu\mu} \la G^2 \ra + (C_5)_{\nu\mu} \la \bar{q}
\sigma_{\alpha\beta} G^{\alpha\beta} q \ra \nnb \\
\ar (C_6)_{\nu\mu} \la \bar{q} \Gamma q \bar{q} \Gamma^\prime q \ra~.
\eea    
The values of the heavy quark condensates are related to the vacuum
expectation value of the gluon operators in the following manner:
\bea
\label{e7315}
\la \bar{Q} Q \ra = - \frac{1}{12 m_Q} \frac{\alpha_s}{\pi} \la G^2 \ra -
\frac{1}{360 m_Q^3} \frac{\alpha_s}{\pi} \la G^3 \ra~,
\eea
where $Q$ is the heavy quark and the heavy quark condensate contributions
are suppressed by inverse of the heavy quark mass, and for this reason we
can safely omit them.

It should be noted here that, in principle, the light quark condensate does 
give contribution to the correlator function, but its contribution becomes
zero after double Borel transformation. Therefore, the only nonperturbative
contribution to the above--mentioned correlator function comes from the gluon
condensate.

As a result, in the lowest order of perturbation theory, the three--point
functions receive contribution from the bare--loop and gluon condensates,
i.e.,
\bea
\label{e7316}
\Pi_i(p_1^2,p_2^2,q^2) = \Pi_i^{per}(p_1^2,p_2^2,q^2) + \Pi_i^{\la G^2 \ra}
(p_1^2,p_2^2,q^2) \frac{\alpha_s}{\pi} \la G^2 \ra~.
\eea
The bare--loop contribution can be obtained using the double
dispersion representation
\bea
\label{e7317}
\Pi_i^{per} = - \frac{1}{(2 \pi)^2} \int \frac{\rho_i^{per} (s,s^\prime,
Q^2)}{(s-p^2) (s^\prime - p^{\prime 2})} ds ds^\prime + \mbox{\rm
subtraction terms}~,
\eea 
in the variable $p^2$ and $p^{\prime 2}$, where $Q^2=-q^2$. The integration
region in Eq. (\ref{e7317}) is determined by the inequalities
\bea
\label{e7318}
-1 \le \frac{2 s s^\prime + (s + s^\prime + Q^2 )(m_c^2 - m_b^2 - s) + 2 s
m_b^2}{\lambda^{1/2} (s,s^\prime,-Q^2) \lambda^{1/2}(m_b^2,m_c^2,s)} \le +1~,
\eea
where $\lambda(a,b,c)=a^2+b^2+c^2-2ab-2ac-2bc$.

The spectral densities $\rho_i^{per} (s,s^\prime,Q^2)$ can be calculated
by using the Gutkovsky rule, i.e., by replacing the propagators with
Dirac--delta functions
\bea
\label{e7319}
\frac{1}{k^2-m^2} \rar -2i\pi \delta(k^2-m^2)~.
\eea
After standard calculations, we get for the spectral densities:
\bea
\label{e7320}
\rho_V \es \frac{2 N_c m_c}{\lambda^{3/2}(s,s^\prime,-Q^2)} (2 s^\prime \Delta_1 -
u \Delta_2)~, \\ \nnb \\
\label{e7321}
\rho_{A_1} \es - \frac{2 N_c m_c}{\lambda^{3/2}(s,s^\prime,-Q^2)}
\Big[ m_c^4 s^\prime + m_c^2 m_b^2 (u - 2 s^\prime) + m_c^2 s^\prime (u-2 s) +
Q^2 (m_b^2 u - s s^\prime - m_b^4) \nnb \\
\ar \frac{1}{2} \lambda(s,s^\prime,-Q^2) \Big]~,\\ \nnb \\
\label{e7322}
\rho_{A_2} \es - \frac{N_c m_c}{\lambda^{3/2}(s,s^\prime,-Q^2)}
\Big[ -(2 s \Delta_2 - u \Delta_1) + B_1 +C - D - E \Big]~,
\eea 
where $N_c=3$ is the color factor, $u=s+s^\prime+Q^2$,
$\Delta_1=s+m_b^2-m_c^2$, $\Delta_2=s^\prime +m_b^2$, and explicit forms of
the functions $B$, $C$, $D$, and $E$ are given in appendix--B. According to
the QCD sum rule philosophy, contributions coming from the excited states are
approximated as bare--loop contribution, starting from some thresholds $s$
and $s^\prime$, in accordance with the quark--hadron duality. Note that we
neglected ${\cal O}(\alpha_s/\pi)$ hard gluon corrections to the bare loop
diagrams, since they are not available yet. However, we expect their
contributions to be about $\sim 10\%$, so that if the accuracy of QCD sum
rules is taken into account, these corrections would not change the results
drastically.  

The next problem is calculation of the gluon condensate contributions to the
correlation functions. The gluon condensate contributions to the three--point
sum rules are described by the diagrams presented in Fig. (1). The
calculations of these diagrams are carried out in the Fock--Schwinger
fixed--point gauge \cite{R7320,R7321,R7322}
\bea
x^\mu A_\mu^a = 0~,\nnb
\eea
where $A_\mu^a$ is the gluon field. In calculating the gluon condensate
contributions, integrals of the following types are encountered:
\bea
\label{e7323}
I_0[a,b,c] \es 
\int \frac{d^4k}{(2 \pi)^4} \frac{1}{\left[ k^2-m_b^2 \right]^a
\left[ (p+k)^2-m_c^2 \right]^b \left[ (p^\prime+k)^2\right]^c}~,
\nnb \\ \nnb \\
I_\mu[a,b,c] \es 
\int \frac{d^4k}{(2 \pi)^4} \frac{k_\mu}{\left[ k^2-m_b^2 \right]^a
\left[ (p+k)^2-m_c^2 \right]^b \left[ (p^\prime+k)^2\right]^c}~,
\nnb \\ \nnb \\
I_{\mu\nu}[a,b,c] \es
\int \frac{d^4k}{(2 \pi)^4} \frac{k_\mu k_\nu}{\left[ k^2-m_b^2 \right]^a  
\left[ (p+k)^2-m_c^2 \right]^b \left[ (p^\prime+k)^2\right]^c}~.
\eea

These integrals can be calculated by continuing to Euclidean space--time and
using Schwinger representation for the Euclidean propagator
\bea
\label{e7324}
\frac{1}{k^2+m^2} = \frac{1}{\Gamma(\alpha)} \int_0^\infty d\alpha \, 
\alpha^{n-1} e^{-\alpha(k^2+m^2)}~,
\eea
which is very suitable for the Borel transformation since 
\bea
\label{e7325}
{\cal B}_{\hat{p}^2} (M^2) e^{-\alpha p^2} = \delta (1/M^2-\alpha)~.
\eea      
Performing integration over loop momentum and over the two parameters which
we have used in the exponential representation of propagators \cite{R7321}, 
and applying double Borel transformations over $p^2$ and $p^{\prime 2}$, 
we get the Borel transformed form of the integrals in Eq. (\ref{e7323})
(see also \cite{R7321})
\bea
\label{e7326}
\hat{I}_0(a,b,c) \es \frac{(-1)^{a+b+c}}{16 \pi^2\,\Gamma(a) \Gamma(b) \Gamma(c)}
(M_1^2)^{2-a-b} (M_2^2)^{2-a-c} \, {\cal U}_0(a+b+c-4,1-c-b)~, \nnb \\ \nnb \\
\hat{I}_\mu(a,b,c) \es \frac{1}{2} \Big[\hat{I}_1(a,b,c) +
\hat{I}_2(a,b,c)\Big] {\cal P}_\mu + \frac{1}{2} \Big[\hat{I}_1(a,b,c) -
\hat{I}_2(a,b,c)\Big] q_\mu~, \nnb \\ \nnb \\ 
\hat{I}_{\mu\nu}(a,b,c) \es \hat{I}_6(a,b,c) g_{\mu\nu} + 
\frac{1}{4} \Big(2 \hat{I}_4 + \hat{I}_3 + \hat{I}_5 \Big) {\cal P}_\mu {\cal P}_\nu  
+ \frac{1}{4} \Big(-\hat{I}_5 + \hat{I}_3 \Big) {\cal P}_\mu q_\nu \nnb \\
\ar \frac{1}{4} \Big(-\hat{I}_5 + \hat{I}_3 \Big) {\cal P}_\nu q_\mu +
\frac{1}{4} \Big(-2 \hat{I}_4 + \hat{I}_3 + \hat{I}_5 \Big) q_\mu q_\nu~,
\eea
where
\bea
\label{e7327}
\hat{I}_1(a,b,c) \es i \frac{(-1)^{a+b+c+1}}{16 \pi^2\,\Gamma(a) \Gamma(b) \Gamma(c)}
(M_1^2)^{2-a-b} (M_2^2)^{3-a-c} \, {\cal U}_0(a+b+c-5,1-c-b)~, \nnb \\ \nnb \\                
\hat{I}_2(a,b,c) \es i \frac{(-1)^{a+b+c+1}}{16 \pi^2\,\Gamma(a) \Gamma(b) \Gamma(c)}
(M_1^2)^{3-a-b} (M_2^2)^{2-a-c} \, {\cal U}_0(a+b+c-5,1-c-b)~, \nnb \\ \nnb \\
\hat{I}_3(a,b,c) \es i \frac{(-1)^{a+b+c}}{32 \pi^2\,\Gamma(a) \Gamma(b) \Gamma(c)}
(M_1^2)^{2-a-b} (M_2^2)^{4-a-c} \, {\cal U}_0(a+b+c-6,1-c-b)~,\nnb \\ \nnb \\
\hat{I}_4(a,b,c) \es i \frac{(-1)^{a+b+c}}{16 \pi^2\,\Gamma(a) \Gamma(b) \Gamma(c)}
(M_1^2)^{3-a-b} (M_2^2)^{3-a-c} \, {\cal U}_0(a+b+c-6,1-c-b)~,\nnb \\ \nnb \\
\hat{I}_5(a,b,c) \es i \frac{(-1)^{a+b+c}}{16 \pi^2\,\Gamma(a) \Gamma(b) \Gamma(c)}
(M_1^2)^{4-a-b} (M_2^2)^{2-a-c} \, {\cal U}_0(a+b+c-6,1-c-b)~,\nnb \\ \nnb \\
\hat{I}_6(a,b,c) \es i \frac{(-1)^{a+b+c+1}}{16 \pi^2\,\Gamma(a) \Gamma(b) \Gamma(c)}
(M_1^2)^{3-a-b} (M_2^2)^{3-a-c} \, {\cal U}_0(a+b+c-6,2-c-b)~,
\eea
where $M_1^2$ and $M_2^2$ are the Borel parameters in the $s$ and $s^\prime$
channel, respectively, and the function ${\cal U}_0(\alpha,\beta)$ is 
defined as
\bea
{\cal U}_0(a,b) = \int_0^\infty dy (y+M_1^2+M_2^2)^a y^b
\,exp\left[ -\frac{B_{-1}}{y} - B_0 - B_1 y \right]~, \nnb
\eea
where 
\bea      
\label{e7328}
B_{-1} \es \frac{m_c^2}{M_1^2} \left[m_c^2 M_2^2 + M_1^2 (m_c^2
+ Q^2) \right] ~, \nnb \\
B_0 \es \frac{1}{M_1^2 M_2^2} \left[ m_b^2 M_1^2 + M_2^2 (m_b^2+m_c^2)
\right] ~, \nnb \\
B_{1} \es \frac{m_b^2}{M_1^2 M_2^2}~.
\eea
Hat in Eqs. (\ref{e7326}) and (\ref{e7327}) means that they are double Borel
transformed form of integrals.
Performing double Borel transformations on the variables $p^2$ and
$p^{\prime 2}$ on the physical parts of the correlator functions and 
bare--loop diagrams and equating two representations of the correlator
functions, we get the sum rules for the form factors $V$, $A_1$ and $A_2$:
\bea
V \es - \frac{(m_b+m_c) (m_{B_c} + m_{B_u^\ast})}{2 f_{B_c} m_{B_c}^2
f_{B_u^\ast} m _{B_u^\ast}} e^{m_{B_c}^2/M_1^2}
e^{m_{B_u^\ast}^2/M_2^2} \nnb \\
\cp \Bigg\{ \frac{1}{4 \pi^2} \int_{m_b^2}^{s_0^\prime} 
ds^\prime \int_{s_L}^{s_0} \rho_V (s,s^\prime,Q^2) e^{-s/M_1^2}
e^{-s^\prime/M_2^2} - i \frac{1}{24 \pi^2} \lla \frac{\alpha_s}{\pi} G^2
\rra C_4^V \Bigg\}~, \nnb \\ \nnb \\
A_1 \es \frac{(m_b+m_c)}{f_{B_c} m_{B_c}^2
f_{B_u^\ast} m _{B_u^\ast}(m_{B_c} + m_{B_u^\ast})} e^{m_{B_c}^2/M_1^2}
e^{m_{B_u^\ast}^2/M_2^2} \nnb \\
\cp \Bigg\{ \frac{1}{4 \pi^2} \int_{m_b^2}^{s_0^\prime}
ds^\prime \int_{s_L}^{s_0} \rho_{A_1} (s,s^\prime,Q^2) e^{-s/M_1^2}
e^{-s^\prime/M_2^2} - i \frac{1}{24 \pi^2} \lla \frac{\alpha_s}{\pi} G^2
\rra C_4^{A_1} \Bigg\}~, \nnb \\ \nnb \\
A_2 \es - \frac{(m_b+m_c)(m_{B_c} + m_{B_u^\ast})}{f_{B_c} m_{B_c}^2
f_{B_u^\ast} m _{B_u^\ast}(m_{B_c}^2 + 3 m_{B_u^\ast}^2+Q^2)} e^{m_{B_c}^2/M_1^2}
e^{m_{B_u^\ast}^2/M_2^2} \nnb \\
\cp \Bigg\{ \frac{1}{4 \pi^2} \int_{m_b^2}^{s_0^\prime}
ds^\prime \int_{s_L}^{s_0} \rho_{A_2} (s,s^\prime,Q^2) e^{-s/M_1^2}
e^{-s^\prime/M_2^2} - i \frac{1}{24 \pi^2} \lla \frac{\alpha_s}{\pi} G^2
\rra C_4^{A_2} \Bigg\}~, \nnb
\eea
where $s_0$ and $s_0^\prime$ are the continuum thresholds in pseudoscalar
$B_c$ and $B_u^\ast$ channels, respectively, and the lower bound integration
limit of $s$ is given as
\bea
\label{e7329}
s_L \es \frac{(m_b^2-Q^2-m_c^2-s^\prime) (m_c^2 s^\prime + m_b^2 Q^2)}
{(m_c^2+Q^2) (m_b^2-s^\prime)}~.
\eea
Explicit expressions of $C_4^V$, $C_4^{A_1}$ and $C_4^{A_2}$ are all
presented in Appendix--C.

At the end of this section we present the dilepton invariant mass
distribution for the $B_c \rar B_u^\ast \ell^+ \ell^-$ decay.
Using the parametrization of the $B_c \rar B_u^\ast$ transition in terms of
form factors and Eq.(\ref{e7301}), the matrix element of the $B_c \rar
B_u^\ast \ell^+ \ell^-$ decay can be written as
\bea
\label{e7330}
{\cal M} = \frac{G \alpha}{4 \sqrt{2} \pi} m_{B_c} \Big[ J_\mu^1 \,
\bar{\ell} \gamma_\mu \ell + J_\mu^2 \, \bar{\ell} \gamma_\mu \gamma_5 \ell
\Big]~,
\eea
where 
\bea
\label{e7331}
J_\mu^1 \es G_1(\hat{s}) \epsilon_{\mu\rho\alpha\beta} \varepsilon^{\ast \rho}
\hat{p}^\alpha \hat{p}^{\prime \beta} - i G_2(\hat{s}) \varepsilon_\mu^{\ast}
+ i G_3 (\hat{s}) (\varepsilon^{\ast} \hat{q}) \hat{{\cal P}}_\mu + i G_4(\hat{s})
(\varepsilon^{\ast} \hat{q}) \hat{q}_\mu~, \nnb \\
J_\mu^2 \es H_1(\hat{s}) \epsilon_{\mu\rho\alpha\beta} \varepsilon^{\ast \rho} 
\hat{p}^\alpha \hat{p}^{\prime \beta} - i H_2(\hat{s}) \varepsilon_\mu^{\ast}     
+ i H_3 (\hat{s}) (\varepsilon^{\ast} \hat{q}) \hat{{\cal P}}_\mu + i H_4(\hat{s})
(\varepsilon^{\ast} \hat{q}) \hat{q}_\mu~,
\eea
where $\hat{{\cal P}}_\mu = {\cal P}_\mu/m_{B_c}$, $\hat{q}_\mu = q_\mu/m_{B_c}$
and $\hat{s} = q^2/m_{B_c}^2$, and
\bea
\label{e7332}
G_1(\hat{s}) \es \frac{2}{1+\hat{r}} C_9^{eff} V(\hat{s}) + \frac{4 \hat{r}_c}{\hat{s}}
C_7^{eff} T_1(\hat{s})~, \nnb \\
G_2(\hat{s}) \es (1+\hat{r})\Bigg[ C_9^{eff} A_1(\hat{s}) + 
\frac{2 \hat{r}_c}{\hat{s}} (1-\hat{r}) C_7^{eff} T_2(\hat{s})\Bigg]~, \nnb \\
G_3(\hat{s}) \es \frac{1}{1-\hat{r}^2} \Bigg\{(1-\hat{r}) C_9^{eff}
A_2(\hat{s}) + 2 \hat{r}_c C_7^{eff} \Bigg[T_3(\hat{s}) +
\frac{1-\hat{r}^2}{\hat{s}} T_2(\hat{s}) \Bigg] \Bigg\}~, \nnb \\
G_4(\hat{s}) \es \frac{1}{\hat{s}} \Bigg\{ C_9^{eff} \Big[ (1+\hat{r})
A_1(\hat{s}) - (1-\hat{r}) A_2(\hat{s}) - 2 \hat{r} A_0(\hat{s}) \Big] -
2 \hat{r}_c C_7^{eff} T_3(\hat{s}) \Bigg\}~, \nnb \\
H_1(\hat{s}) \es \frac{2}{1+\hat{r}} C_{10} V(\hat{s})~, \nnb \\
H_2(\hat{s}) \es (1+\hat{r}) C_{10} A_1(\hat{s})~, \nnb \\
H_3(\hat{s}) \es \frac{1}{1+\hat{r}} C_{10} A_2(\hat{s})~, \nnb \\
H_4(\hat{s}) \es \frac{1}{\hat{s}} C_{10} \Big[ (1+\hat{r}) A_1(\hat{s}) - 
(1-\hat{r}) A_2(\hat{s}) - 2 \hat{r} A_0(\hat{s}) \Big]~, 
\eea
where $\hat{r} = m_{B_u^\ast}/m_{B_c}$ and $\hat{r}_c = m_c/m_{B_c}$.

Using Eq. (\ref{e7330}), the dilepton invariant mass distribution takes the
following form:
\bea
\label{e7333}
\frac{d \Gamma}{d\hat{s}} \es \frac{G^2 \alpha_s^2}{2^{10} \pi^5} m_{B_c}^5
\sqrt{\lambda} v \Bigg\{
\frac{1}{3} \hat{s} \lambda \Bigg(1 +2 \frac{\hat{r}_\ell^2 }{\hat{s}} \Bigg)
\vel G_1 \ver^2 + \frac{1}{3} \hat{s} \lambda v^2 \vel H_1 \ver^2 \nnb \\
\ar \frac{1}{4 \hat{r}^2} \Bigg[ \Bigg( \lambda - \frac{\lambda v^2}{3} + 8
\hat{r}^2 (\hat{s} + 2 \hat{r}_\ell^2) \Bigg) \vel G_2 \ver^2 +
\Bigg( \frac{\lambda}{3} (3-v^2) + 8 \hat{r}^2 \hat{s} v^2 \Bigg) \vel H_2 \ver^2 
\Bigg]\nnb \\
\ar \frac{\lambda}{4 \hat{r}^2} \Bigg[\frac{\lambda}{3} (3-v^2) \vel G_3 \ver^2 +
\Bigg( \frac{\lambda}{3} (3-v^2) + 4 \hat{r}_\ell^2 (2 + 2 \hat{r}^2 -
\hat{s}) \Bigg) \vel H_3 \ver^2\Bigg] \nnb \\
\ek \frac{1}{2 \hat{r}^2} \Bigg[
\frac{\lambda}{3} (3-v^2) (1-\hat{r}^2 - \hat{s}) \mbox{\rm Re}[G_2 G_3^\ast] +
\Bigg( \frac{\lambda}{3} (3-v^2)
(1-\hat{r}^2 - \hat{s}) + 4 \hat{r}_\ell^2 \lambda \Bigg) 
\mbox{\rm Re}[H_2 H_3^\ast]\Bigg] \nnb \\
\ek 2 \lambda \frac{\hat{r}_\ell^2}{\hat{r}^2} \Big( \mbox{\rm Re}[H_2 H_4^\ast]
- (1-\hat{r}^2) \mbox{\rm Re}[H_3 H_4^\ast] \Big) +
\frac{\hat{r}_\ell^2}{\hat{r}^2} \hat{s} \lambda \vel H_4 \ver^2 \Bigg\}~,  
\eea
where
\bea
\lambda = \lambda(1,\hat{r}^2,\hat{s}),~v = \sqrt{1-\frac{4
m_\ell^2}{q^2}}~\mbox{\rm is the lepton velocity},~ r_\ell^2 =
\frac{m_\ell^2}{m_{B_c}^2}~.\nnb
\eea

\section{Numerical analysis}

In this section we present our numerical calculation of the form factors
$A_1$, $A_2$, $A_0$, $V$, $T_1$, $T_2$ and $T_3$. The values of the input
parameters appearing in the sum rules for the form factors are:
$m_{B_c}=6.4~GeV$, $m_{B_u^\ast}=5.325~GeV$, $f_{B_c} =0.385~GeV$
\cite{R7324},
$f_{B_u^\ast}=160~MeV$ \cite{R7325}, 
$m_b=4.8~GeV$, $m_c(\mu=m_c)=1.26~GeV$, 
$\la (\alpha_s/\pi) G^2 \ra=0.012~GeV^4$\cite{R7319}. The decay constants of
$B_c$ and $B_u^\ast$ mesons are determined from the two--point QCD sum
rules. As has already been noted, in  bare--loop calculations we neglect 
${\cal O}(\alpha_s/\pi)$ corrections. For consistency, these corrections are also
neglected in the calculations for the leptonic decay constants. 

The parameters $s_0$ and $s_0^\prime$, which are the continuum thresholds 
of $B_c$ and $B_u^\ast$ mesons, respectively, are also determined from the 
two--point QCD sum rules, and they are taken to be $s_0=50~GeV^2$ and 
$s_0^\prime=35~GeV^2$. These continuum thresholds are determined from the
conditions that guarantees the sum rules to have the best stability 
in the allowed $M_1^2$ and $M_2^2$ region.

The Borel parameters $M_1^2$ and $M_2^2$ in the sum rules are auxiliary
parameters and physical quantities should be independent of them. Therefore
it is necessary to look for a working regions of these parameters where
physical results exhibit best stability. The working regions of $M_1^2$ and
$M_2^2$ are determined by requiring that the continuum and higher states
contributions are effectively suppressed, which ensures that the results do
not sensitively depend on such excited states. We require also that the
contribution gluon condensate is not too large, which guarantees that the
contributions of higher dimensional operators are small. Our analysis
verifies that the working regions $20~GeV^2 \le M_1^2 \le 40~GeV^2$ and
$10~GeV^2 \le M_2^2 \le 15~GeV^2$ of the Borel parameters satisfy both of
the above--mentioned requirements.

In these regions of $M_1^2$ and $M_2^2$, the gluon condensate contribution
constitutes approximately $5\%$, and higher state contributions constitute 
at most $30\%$ of the total result.

Our numerical results for the form factors at $q^2=0$ are:
\bea
\label{e7334}
V(0)   \es  0.09 \pm 0.02~, \nnb \\
A_1(0) \es -0.17 \pm 0.03~, \nnb \\
A_2(0) \es  1.10 \pm 0.20~, \nnb \\
T_1(0) \es  T_2(0) = - A_0(0) = 0.23 \pm 0.04~,
\eea 
and the value of the form factor $T_3(0)$ can easily be obtained from Eq.
(\ref{e7307}).
        
The errors are estimated by the variation of the Borel parameters $M_1^2$
and $M_2^2$, the variation of the continuum thresholds $s_0$ and
$s_0^\prime$, the variation of $b$ and $c$ quark masses and leptonic decay
constants $f_{B_c}$ and $f_{B_u^\ast}$. The main uncertainty comes from
the thresholds and the decay constants, which is about $\sim 20\%$ of the
central value, while the other uncertainties are small, constituting a few
percent. Note that all the uncertainties are added quadratically.
Here we would like to make the following cautionary note. It is well known
that for heavy quarkonia, where the quark velocities are small, the
$\alpha_s/v$ corrections caused by the Coulomb--like interaction of quarks,
are essential, where $v$ is the quark velocity. In our case, we have two
expansion parameters, $\alpha_s/v_1$ and $\alpha_s/v_2$, where $v_1$ and
$v_2$ are the relative velocities of quarks $(b\bar{c})$ and $(\bar{b}u)$
(for massless $u$ quark $v_2=1$). When these corrections are taken into
account the value of the form factors at $Q^2=0$ are twice as greater.   
In further numerical analysis we will omit the dependence of the form
factors on $q^2$, which gives small contributions to the overall result.
Indeed, the maximum value of $q^2$ in the decay under consideration is
about $(m_{B_c} - m_{B_u^\ast})^2 \sim 1~GeV^2$ and assuming that the simple
pole model correctly describes the $q^2$ dependence of the form factors, it
is easy to see that $q^2/m_{B_c}^2$ is about $1/36$; i.e., the
results could be changed maximally about $3\%$. 
 
Integrating Eq. (\ref{e7332}) over $q^2$ in the whole physical region and
using the total mean life time $\tau \simeq 0.46~ps$ of $B_c$ meson
\cite{R7326}, the branching ratio of the $B_c \rar B_u^\ast \ell^+ \ell^-$ 
decay is
\bea
{\cal B}(B_c \rar B_u^\ast \ell^+ \ell^-)  = \left\{ \begin{array}{l}
1.3\times 10^{-9}~,\\ \\
3.9\times 10^{-7}~. \end{array} \right.  \nnb
\eea
where the upper value corresponds to the case when only short distance
contributions are taken into account, and the lower one corresponds when
short and long distance contributions due to the $\rho$ and $\omega$
resonances are taken into account.
 
It follows from this result that, the $B_c \rar B_u^\ast \ell^+ \ell^-$
decay with the above--presented width can be measurable at LHC.

In conclusion, we calculate the $B_c \rar B_u^\ast$ transition form factors
$V(q^2)$, $A_1(q^2)$, $A_2(q^2)$ and $T_1(q^2)$ in the framework of QCD sum 
rule. Our calculations show that this rare, semileptonic decay can be 
measurable at LHC. Using the results of the form factors calculated in the 
present work, we estimate the branching ratio of the 
$B_c \rar B_u^\ast \ell^+ \ell^-$ decay.

\section*{Acknowledgments}

One of the authors (T.M.A) is grateful to S. Fajfer for the useful
discussion and T\"{U}B\.{I}TAK for partially support of this work under
project 105T131.

\newpage

\appendix

\begin{center}
{\Large{\bf Appendix--A}}
\end{center}


\setcounter{equation}{0}   
\renewcommand{\theequation}{A.\arabic{equation}}
\setcounter{section}{0}
\setcounter{table}{0}

\section*{}
In this appendix we present the expressions of the functions $F_1$, $F_2$
$C^{box}$ and $C^Z$, which enter to the expressions of the Wilson
coefficients $C_7(m_W)$, $C_9(m_W)$ and $C_{10}(m_W)$:
\bea
C^{box}(x_i) \es \frac{3}{8} \Bigg[ - \frac{1}{x_i-1} + \frac{x_i \ln
x_i}{(x_i-1)^2} \Bigg]~,\nnb \\ \nnb \\
C^Z(x_i) \es \frac{x_i}{4} - \frac{3}{8} \frac{1}{x_i-1} + \frac{3}{8}
\frac{2 x_i^2-x_i}{8(x_i-1)^2} \ln x_i~, \nnb \\ \nnb \\
F_1(x_i) \es Q_d \Bigg\{ \Bigg[ \frac{1}{12} \frac{1}{x_i-1} + \frac{13}{12}
\frac{1}{(x_i-1)^2} - \frac{1}{2(x_i-1)^3} \Bigg] x_i \nnb \\
\ar \Bigg[\frac{2}{3} \frac{1}{x_i-1}
+ \Bigg( \frac{2}{3} \frac{1}{(x_i-1)^2} -
\frac{5}{6} \frac{1}{(x_i-1)^3} + \frac{1}{2} \frac{1}{(x_i-1)^4} \Bigg) x_i
\Bigg] \ln x_i \Bigg\}~ \nnb \\
\ek \Bigg[\frac{7}{3} \frac{1}{x_i-1} + \frac{13}{12} \frac{1}{(x_i-1)^2}
- \frac{1}{2} \frac{1}{(x_i-1)^3}\Big] x_i \nnb \\
\ek \Bigg[\frac{1}{6} \frac{1}{x_i-1} - \frac{35}{12} \frac{1}{(x_i-1)^2} -
\frac{5}{6} \frac{1}{(x_i-1)^3} + \frac{1}{2} \frac{1}{(x_i-1)^4} \Bigg] x_i
\ln x_i~, \nnb \\ \nnb \\
F_2(x_i) \es - Q_d \Bigg\{ \Bigg[- \frac{1}{4} \frac{1}{x_i-1} + \frac{3}{4}
\frac{1}{(x_i-1)^2} + \frac{3}{2} \frac{1}{(x_i-1)^3} \Bigg] -
\frac{3}{2} \frac{x_i^2 \ln x_i}{(x_i-1)^4} \Bigg\} \nnb \\
\ar \Bigg[ \frac{1}{2} \frac{1}{x_i-1} + \frac{9}{4} \frac{1}{(x_i-1)^2} +
\frac{3}{2} \frac{1}{(x_i-1)^3} \Bigg] x_i - \frac{3}{2} \frac{x_i^3 \ln
x_i}{2 (x_i-1)^4}~,
\eea
where $x_i = m_i^2/m_W^2$, and $Q_d$ is the down quark charge. Note that ,
in these expressions, we omit the gauge dependent terms $\gamma(\xi,x_i)$
\cite{R7316}, because these terms are canceled out in the combinations
$C^{box}(x_i) + C^Z(x_i)$ and $F_1(x_i) + C^Z(x_i)$.

\newpage
  
\begin{center}
{\Large{\bf Appendix--B}}
\end{center}

\setcounter{equation}{0}   
\renewcommand{\theequation}{B.\arabic{equation}}
\setcounter{section}{0}
\setcounter{table}{0}

In this appendix we present the expressions of the functions $B_1$, $C$,
$D$, and $E$, which appear in the calculations of the spectral density
$\rho_{A_2}$ (see eq. (\ref{e7322})).

\bea
B_1 \es \frac{1}{\lambda(s,s^\prime,-Q^2)} I_0 \Big\{
m_b^2 \Big[ (Q^2 + s)^2 - 2 s^\prime (2 Q^2 + s) +s^{\prime 2}\Big] \nnb \\
\ar 2 m_b^2 s^\prime \Big[Q^4 - (s-s^\prime)^2 + 3 m_c^2 (u-2 s^\prime) +
Q^2 u \Big] \nnb \\
\ar s^{\prime 2} \Big[ 6 m_c^4 + Q^4 + (s-s^\prime)^2 + 6 m_c^2 (u-2 s) +
2 Q^2 (s-2 s^\prime) \Big] \Big\}~, \nnb \\ \nnb \\ 
C \es D = \frac{1}{\lambda^2(s,s^\prime,-Q^2)} I_0 \Big\{ m_b^2 (u-s^\prime) 
\Big[ -Q^4 + m_b^2 (2 Q^2 -s) + s^2 \Big] \nnb \\
\ar s^\prime \Big[ s (2 Q^2 -s ) (Q^2 + s) + m_b^4 (Q^2 + 2s) - m_b^2 (Q^4 + 6
Q^2 s + s^2) \Big] \nnb \\
\ar s^{\prime^2} \Big[ m_b^4 + m_b^2 (s-Q^2) -s (Q^2 + 2s)\Big] +
s^{\prime^3}
(m_b^2-s) - 3 m_c^4 s^\prime u \nnb \\
\ar 2 m_c^2 \Big[ - m_b^2 \Big( (u-s^\prime)^2 + s^\prime (s-Q^2) -2
s^\prime \Big) - s^\prime \Big(Q^2 (u-2 s) - 2 s^2 + s^\prime u\Big) \Big]
\Big\}~, \nnb \\ \nnb \\
E \es \frac{1}{\lambda^{3/2}(s,s^\prime,-Q^2)} I_0 \Big\{ s^2 \Big[ m_c^4 ( u + 6
s s^\prime ) + 2 m_b^2 s \Big[ Q^4 - (s-s^\prime)^2 + Q^2 u \Big] \nnb \\
\ar m_b^4 \Big[ Q^4 + (s-s^\prime)^2 + 2 Q^2 (-2 s + s^\prime) \Big] +
2 m_c^2 \Big[-m_b^3 \Big( q^2 (u-2s) -2 s^2 + s^\prime u\Big) \nnb \\
\ar s \Big( -(u-s^\prime)^2 + s^\prime (-s +Q^2) + 2 s^{\prime^2} \Big) \Big]
\Big\}~,
\eea
where 
\bea
I_0 = \frac{1}{4 \lambda^{1/2} (s,s^\prime,-Q^2)}~.\nnb
\eea

\newpage

\appendix

\begin{center}
{\Large{\bf Appendix--C}}
\end{center}


\setcounter{equation}{0}   
\renewcommand{\theequation}{C.\arabic{equation}}
\setcounter{section}{0}
\setcounter{table}{0}

\section*{}
In this appendix we give the explicit expressions of the coefficients of the
gluon condensate which enter to the sum rules for the form factors $V$,
$A_1$ and $A_2$, respectively.
\bea
C_4^V \es
192 m_c \hat{I}_1(1, 3, 1) + 192 m_c^3 \hat{I}_1(1, 4, 1)
-32 m_c \hat{I}_1(2, 1, 2) \nnb \\
\ar 64 m_c \hat{I}_1(1, 1, 3) + 32 m_c \hat{I}_1(2, 1, 2) + 128 m_c m_b^2 \hat{I}_1(2, 1, 3) - 
 128 m_c 
\hat{I}_1^{[0,1]}(2, 1, 3) \nnb \\
\ar 64 m_c \hat{I}_1(3, 1, 1) + 192 m_c m_b^2 \hat{I}_1(3, 1, 2) - 
 128 m_c 
\hat{I}_1^{[0,1]}(3, 1, 2) + 64 m_c m_b^4 \hat{I}_1(3, 1, 3) \nnb \\
\ek 128 m_c m_b^2 
\hat{I}_1^{[0,1]}(3, 1, 3) + 64 m_c 
\hat{I}_1^{[0,2]}(3, 1, 3)
+ 32 m_c \hat{I}_1(1, 2, 2) + 32 m_c \hat{I}_1(2, 2, 1) \nnb \\
\ar 32 m_c m_b^2 \hat{I}_1(2, 2, 2) - 
 32 m_c 
\hat{I}_1^{[0,1]}(2, 2, 2) 
-32 m_c \hat{I}_1(1, 2, 2) 
+ 192 m_c m_b^2 \hat{I}_1(4, 1, 1) \nnb \\
\ar 96 m_c \hat{I}_1(2, 2, 1)~,
\eea
\bea
C_4^{A_1} \es 
96 m_c m_b^2 \hat{I}_0(1, 3, 1) - 96 m_c 
\hat{I}_0^{[0,1]}(1, 3, 1) + 96 m_c^3 m_b^2 \hat{I}_0(1, 4, 1) \nnb \\
\ek 96 m_c^3 
\hat{I}_0^{[0,1]}(1, 4, 1) - 384 m_c \hat{I}_6(1, 3, 1) - 384 m_c^3 \hat{I}_6(1, 4, 1)
-16 m_c \hat{I}_0(1, 1, 2) \nnb \\
\ek 16 m_c \hat{I}_0(2, 1, 1) - 16 m_c m_b^2 \hat{I}_0(2, 1, 2) + 
 16 m_c 
\hat{I}_0^{[0,1]}(2, 1, 2) - 64 m_c \hat{I}_6(2, 1, 2) \nnb \\
\ek 16 m_c \hat{I}_0(1, 1, 2) + 96 m_c m_b^2 \hat{I}_0(1, 1, 3) - 96 m_c 
\hat{I}_0^{[0,1]}(1, 1, 3) - 
 16 m_c \hat{I}_0(2, 1, 1) \nnb \\
\ar 48 m_c m_b^2 \hat{I}_0(2, 1, 2) - 80 m_c 
\hat{I}_0^{[0,1]}(2, 1, 2) + 
 96 m_c m_b^4 \hat{I}_0(2, 1, 3) - 192 m_c m_b^2 
\hat{I}_0^{[0,1]}(2, 1, 3) \nnb \\
\ar 96 m_c 
\hat{I}_0^{[0,2]}(2, 1, 3) + 64 m_c m_b^2 \hat{I}_0(3, 1, 1) - 96 m_c 
\hat{I}_0^{[0,1]}(3, 1, 1) + 
 64 m_c m_b^4 \hat{I}_0(3, 1, 2) \nnb \\
\ek 160 m_c m_b^2 
\hat{I}_0^{[0,1]}(3, 1, 2) + 
 96 m_c 
\hat{I}_0^{[0,2]}(3, 1, 2) + 32 m_c m_b^6 \hat{I}_0(3, 1, 3) - 
 96 m_c m_b^4 
\hat{I}_0^{[0,1]}(3, 1, 3) \nnb \\
\ar 96 m_c m_b^2 
\hat{I}_0^{[0,2]}(3, 1, 3) - 
 32 m_c 
\hat{I}_0^{[0,3]}(3, 1, 3) + 64 m_c \hat{I}_6(2, 1, 2) - 128 m_c m_b^2 \hat{I}_6(2, 1, 3) \nnb \\
\ar 128 m_c 
\hat{I}_6^{[0,1]}(2, 1, 3) - 256 m_c m_b^2 \hat{I}_6(3, 1, 2) + 
 128 m_c 
\hat{I}_6^{[0,1]}(3, 1, 2) - 128 m_c m_b^4 \hat{I}_6(3, 1, 3) \nnb \\
\ar 256 m_c m_b^2 
\hat{I}_6^{[0,1]}(3, 1, 3) - 128 m_c 
\hat{I}_6^{[0,1]}(3, 1, 3) 
+ 32 m_c m_b^2 \hat{I}_0(1, 2, 2) - 32 m_c 
\hat{I}_0^{[0,1]}(1, 2, 2) \nnb \\
\ek 32 m_c 
\hat{I}_0^{[0,1]}(2, 2, 1) + 
 16 m_c m_b^4 \hat{I}_0(2, 2, 2) - 32 m_c m_b^2 
\hat{I}_0^{[0,1]}(2, 2, 2) + 
 16 m_c 
\hat{I}_0^{[0,2]}(2, 2, 2) \nnb \\
\ek 128 m_c \hat{I}_6(1, 2, 2) + 64 m_c m_b^2 \hat{I}_6(2, 2, 2) + 
 64 m_c 
\hat{I}_6^{[0,1]}(2, 2, 2)
-16 m_c \hat{I}_0(1, 2, 1) \nnb \\
\ek 16 m_c m_b^2 \hat{I}_0(1, 2, 2) + 16 m_c 
\hat{I}_0^{[0,1]}(1, 2, 2) - 
 64 m_c \hat{I}_6(1, 2, 2)
+ 96 m_c m_b^2 \hat{I}_0(3, 1, 1) \nnb \\
\ar 96 m_c m_b^4 \hat{I}_0(4, 1, 1) - 
 96 m_c m_b^2 
\hat{I}_0^{[0,1]}(4, 1, 1) - 384 m_c m_b^2 \hat{I}_6(4, 1, 1)
+48 m_c \hat{I}_0(1, 2, 1) \nnb \\
\ar 48 m_c m_b^2 \hat{I}_0(2, 2, 1) - 48 m_c 
\hat{I}_0^{[0,1]}(2, 2, 1) - 
 192 m_c \hat{I}_6(2, 2, 1)~,
\eea
\bea
C_4^{A_2} \es 
96 m_c \hat{I}_2(1, 3, 1) + 96 m_c^3 \hat{I}_2(1, 4, 1) - 96 m_c \hat{I}_3(1, 3, 1) - 
 96 m_c^3 \hat{I}_3(1, 4, 1) \nnb \\
\ar 96 m_c \hat{I}_5(1, 3, 1) + 96 m_c^3 \hat{I}_5(1, 4, 1)
+ 16 m_c \hat{I}_2(2, 1, 2) - 16 m_c \hat{I}_3(2, 1, 2) + 16 m_c \hat{I}_5(2, 1, 2)
\nnb \\
\ek 32 m_c \hat{I}_0(1, 1, 3) - 32 m_c \hat{I}_0(2, 1, 2) - 32 m_c m_b^2 \hat{I}_0(2, 1, 3) + 
 32 m_c 
\hat{I}_0^{[0,1]}(2, 1, 3) - 32 m_c \hat{I}_2(1, 1, 3) \nnb \\
\ek 16 m_c \hat{I}_2(2, 1, 2) + 
 32 m_c \hat{I}_2(3, 1, 1) + 96 m_c m_b^2 \hat{I}_2(3, 1, 2) - 64 m_c 
\hat{I}_2^{[0,1]}(3, 1, 2) \nnb \\
\ar 32 m_c m_b^4 \hat{I}_2(3, 1, 3) - 64 m_c m_b^2 
\hat{I}_2^{[0,1]}(3, 1, 3) + 
 32 m_c 
\hat{I}_2^{[0,2]}(3, 1, 3) + 16 m_c \hat{I}_3(2, 1, 2) \nnb \\
\ek 32 m_c m_b^2 \hat{I}_3(2, 1, 3) + 
 32 m_c 
\hat{I}_3^{[0,1]}(2, 1, 3) - 64 m_c m_b^2 \hat{I}_3(3, 1, 2) + 32 m_c 
\hat{I}_3^{[0,1]}(3, 1, 2) \nnb \\
\ek 32 m_c m_b^4 \hat{I}_3(3, 1, 3) + 64 m_c m_b^2 
\hat{I}_3^{[0,1]}(3, 1, 3) - 
 32 m_c 
\hat{I}_3^{[0,2]}(3, 1, 3) - 16 m_c \hat{I}_5(2, 1, 2) \nnb \\
\ar 32 m_c m_b^2 \hat{I}_5(2, 1, 3) - 
 32 m_c 
\hat{I}_5^{[0,1]}(2, 1, 3) + 64 m_c m_b^2 \hat{I}_5(3, 1, 2) - 32 m_c 
\hat{I}_5^{[0,1]}(3, 1, 2) \nnb \\
\ar 32 m_c m_b^4 \hat{I}_5(3, 1, 3) - 64 m_c m_b^2 
\hat{I}_5^{[0,1]}(3, 1, 3) + 
 32 m_c 
\hat{I}_5^{[0,2]}(3, 1, 3)
+ 16 m_c \hat{I}_0(1, 2, 2) \nnb \\
\ek 32 m_c m_b^2 \hat{I}_0(2, 2, 2) + 48 m_c \hat{I}_2(1, 2, 2) + 
 16 m_c \hat{I}_2(2, 2, 1) - 48 m_c m_b^2 \hat{I}_2(2, 2, 2) \nnb \\
\ek 16 m_c 
\hat{I}_2^{[0,1]}(2, 2, 2) - 
 32 m_c \hat{I}_3(1, 2, 2) + 16 m_c m_b^2 \hat{I}_3(2, 2, 2) + 16 m_c 
\hat{I}_3^{[0,1]}(2, 2, 2) \nnb \\
\ar 32 m_c \hat{I}_5(1, 2, 2) - 16 m_c m_b^2 \hat{I}_5(2, 2, 2) - 16 m_c 
\hat{I}_5^{[0,1]}(2, 2, 2)
-32 m_c \hat{I}_1(1, 2, 2) \nnb \\
\ar 16 m_c \hat{I}_2(1, 2, 2) - 16 m_c \hat{I}_3(1, 2, 2) + 
16 m_c \hat{I}_5(1, 2, 2)
+ 96 m_c m_b^2 \hat{I}_2(4, 1, 1) \nnb \\
\ek 96 m_c m_b^2 \hat{I}_3(4, 1, 1) + 96 m_c m_b^2 \hat{I}_5(4, 1, 1)
+ 48 m_c \hat{I}_2(2, 2, 1) - 48 m_c \hat{I}_3(2, 2, 1) \nnb \\
\ar 48 m_c \hat{I}_5(2, 2, 1)~,
\eea
 
where
\bea
\hat{I}_n^{[i,j]} (a,b,c) = \ga M_1^2 \dr^i \ga M_2^2 \dr^j \frac{d^i}{d\ga
M_1^2 \dr^i} \frac{d^j}{d\ga M_2^2 \dr^j} \left[\ga M_1^2 \dr^i \ga M_2^2
\dr^j \hat{I}_n(a,b,c) \right]~. \nnb
\eea


\newpage

\begin{thebibliography}{99}

\bibitem{R7301}
 F. Abe {\it et al.}, CDF Collaboration,
 Phys. Rev. D {\bf 58}, 112004 (1998);
 Phys. Rev. Lett. {\bf 81}, 2432 (1998).

\bibitem{R7302}
 A. Abulencia {\it et al.}, CDF Collaboration,
 prep: hep--ex/0505076 (2005).

\bibitem{R7303}
 V. V. Kiselev, 
 prep: hep--ph/0308214 (2003).

\bibitem{R7304}
 I. P. Gouz, V. V. Kiselev, A. K. Likhoded, V. I. Romanovsky and
 O. P. Yushchenko, 
 Phys. Atom. Nucl. {\bf 67} (2004) 1559.      

\bibitem{R7305}
 Y. Grossman,
 Int. J. Mod. Phys. A {\bf 19}, 907 (2004).

\bibitem{R7306}
 G. Isidori, 
 prep: hep--ph/0401079 (2004).

\bibitem{R7307}
 A. J. Buras, 
 prep: hep--ph/0402191 (2004).

\bibitem{R7308}
 T. Ohl {\it et al.},
 Nucl. Phys. B {\bf 403}, 605 (1993);
 J. Donoghue, E. Golowich, B. Holstein and J. trampetic,
 Phys. Rev. D {\bf 33}, 179 (1986);
 A. A. Petrov,
 Phys. Rev. D {\bf 56}, 1685 (1997). 

\bibitem{R7309}
 E. Golowich, and A. A. Petrov,
 Phys. Lett. B {\bf 427}, 172 (1998);
 A. Falk, Y. Grossman, Z. ligeti and A. A. Petrov, 
 Phys. Rev. D {\bf 69}, 114021 (2004).

\bibitem{R7310}
 G. Burdman, E. Golowich, J. Hewett and S. Pakvasa,
 Phys. Rev. D {\bf 52}, 6383 (1995).

\bibitem{R7311}
 C. Greub {\it et al.},
 Phys. Lett. B {\bf 382}, 415 (1996).

\bibitem{R7312}
 B. Bajc, S. Fajfer and R. J. Oakes,
 Phys. Rev. D {\bf 54}, 5883 (1996);
 S. Fajfer, S. Prelovsek and P. Singer,
 Eur.  Phys. J. C {\bf 6}, 471 (1999).

\bibitem{R7313}         
 S. Fajfer, S. Prelovsek and P. Singer,
 Phys. Rev. D {\bf 59}, 114003 (1999).

\bibitem{R7314}
 T. M. Aliev, M. Savc{\i},
 Phys. Lett. B {\bf 480}, 97 (2000).

\bibitem{R7315}
 G. Burdman, E. Golowich, J. Hewett and S. Pakvasa,
 Phys. Rev. D {\bf 66}, 014009 (2002).

\bibitem{R7316}  
 T. Inami and C. S. Lim,
 Prog. Theor. Phys. {\bf 65}, 197 (1981);
 Errata, {\bf 65}, 1172 (1981).

\bibitem{R7317}  
 C. S. Lim, T. Morozumi and A. I. Sanda,
 Phys. Lett. B {\bf 218}, 343 (1989);
 N. G. Deshpande, J. Trampetic and K. Panrose,
 Phys. Rev. D {\bf 39}, 1461 (1989).

\bibitem{R7318}
 M. Soares,
 Phys. Rev. D {\bf 54}, 6837 (1996).

\bibitem{R7319}  
 M. A. Shifman, A. I. Vainshtein and V. I. Zakharov,
 Nucl. Phys. B {\bf 147}, 385 (1979).

\bibitem{R7320}  
 V. A. Fock,
 Sov. Phys. {\bf 12}, 404 (1937).

\bibitem{R7321}                
 J. Schwinger,
 Phys. Rev. {\bf 82}, 664 (1951).

\bibitem{R7322}
 V. Smilga,
 Sov. J. Nucl. Phys. {\bf 35}, 215 (1982).

\bibitem{R7323}
 V. V. Kiselev, A. K. Likhoded and A. I. Onishchenko,
 Nucl. Phys. B {\bf 569}, 473 (2000).

\bibitem{R7324}   
 V. V. Kiselev, A. V. Tkabladze,
 Sov. J. Nucl. Phys. {\bf 50}, 1063 (1989);
 T. M. Aliev, O. Y{\i}lmaz,
 Nuove Cimento A {\bf 105}, 827 (1992).

\bibitem{R7325}    
 V. M. Belyaev, V. M. Braun, A. Khodjamirian and R. R\"{u}ckl,
 Phys. Rev. D {\bf 51}, 6177 (1995).

\bibitem{R7326}
 A. Abulencia, CDF Collaboration,
 prep. hep--ex/0603027 (2006). 


\end{thebibliography}
\end{document}